\documentstyle[a4,11pt]{article}
\begin{document}  
\begin{center}
Notes on Riemann geometry and integrable systems. Part IV.
\end{center}

\begin{center}
Kur. Myrzakul\footnote{Permanent address:
Institute of Mathematics,  Alma-Ata, Kazakhstan } and 
R.Myrzakulov\footnote{Permanent address: Institute of Physics and Technology,
480082,  Alma-Ata-82,  Kazakhstan. E-mail: cnlpmyra@satsun.sci.kz}
\end{center}

\begin{center}
Department of Mathematics, Bilkent University, 06533 Ankara, Turkey 
\end{center}

\begin{abstract}
The connection between multidimensional soliton equations and 
three-dimensional Riemann space is discussed.
\end{abstract}

\tableofcontents
\section{Introduction}

As well-known the relation between geometry and (1+1)-dimensional 
solitonic equations is described (the geometrical part) in terms of 
two-dimensional surfaces 
(and/or curves that is equivalent in 1+1) in a 3-dimensional $M^{3}$ 
(Euclidean or pseudo-Euclidean) spaces [1-24]. In 2+1 dimensions the 
situation is more complicated [25-30, 36-38, 40]. 
Usually, in this case uses the same approach, i.e., two-dimensional 
surfaces arbitrarily embedded in 3-dimensional space $M^{3}$
but considering also the motion (deformation) of surfaces. Of course, 
no doubt that it is correct but in some sense is the artificial 
construction.  Sometimes we think  that in 2+1 dimensions 
may be the more natural geometry  
(at least than the above mentioned surfaces approach) is 3-dimensional 
Riemann spaces (see, e.g., refs.[26-29, 46-48]).   This line we realized in 
our notes [49-51]. 

This note is a sequel to the 
preceding notes [49-51]. 
Our main goal in this note  is to discuss the  interrelation between 
the 
intrinsic geometry of three-dimensional Riemann spaces and integrable 
systems, in particular, spin systems in 2+1 dimensions without entering 
into details (for details see e.g. refs.[49-51]). We will do 
this without any reference to the enveloping spaces (see, also, ref. 
[26]). We will show that 
three-dimensional Riemann spaces is important not only in the general 
relativity but is also important in the theory of multidimensional 
integrable systems.  As well known, the theories of multidimensional 
integrable systems  
and three-dimensional Riemann spaces are not as complete as their 
counterparts, respectively, integrable systems in 1+1 dimensions and 
two-dimensional surfaces. So, the study of the connections between 
integrable systems in 2+1 dimensions  and three-dimensional Riemann 
space is one of actual problems of modern mathematical physics.

\section{Three-dimensional Riemann  space}
Let $V^{3}$ be the space endowed with the affine connection. In this 
space we introduce twoo systems of coordinates: $(x)=(x^{1}, x^{2}, 
x^{3})$ and $(y)=(y^{1}, y^{2}, y^{3})$. It is well known from the 
classical differential geometry that these coordinate systems are 
connected by the following set of equations of second order (remark: 
for convenience, in [49-51] and this note we will use the unified 
common numerations for formulas)
$$ \frac{\partial^{2}y_{k}}{\partial 
x_{i}\partial 
x_{j}}= \Gamma^{l}_{ij}(x)\frac{\partial y_{k}}{\partial x_{l}}-
\Gamma^{k}_{lm}(y)\frac{\partial y_{l} \partial y_{m}}{\partial 
x_{i}\partial x_{j}}  \eqno(155)
$$

We mention that in this case the curvature tensor
$$
R^{i}_{klm}=[\frac{\partial \Gamma_{m}}{\partial x^{l}}
-\frac{\partial \Gamma_{l}}{\partial x^{n}}
+\Gamma_{l}\Gamma_{m}-
\Gamma_{m}\Gamma_{l}]^{i}_{k}
\eqno(156)
$$
has only three independent components.           
Let the space $V^{3}$ is flat and let   the metric tensor 
in $V^{3}$ in the 
coordinate system $(y)$ is diagonal:
$$
ds^{2}=\sum^{3}_{i,j=1}\mu_{ij}dy^{i}dy^{j}   \eqno(157)
$$
with $\mu_{ij}=\pm 1$. Hence follows that  the cooresponding connection 
and   Riemann's curvature tensor are equal to zero: 
$$
\Gamma^{i}_{jk}(y)=0,  \quad R^{i}_{klm}(y)=0
\eqno(158)
$$

Let for the coordinate system $(x)$ the metric  has the form
$$
ds^{2}=\sum^{3}_{i,j=1}g_{ij}dx_{i}dx_{j}.   \eqno(159)
$$                        
As the curvature tensor has the law of transformation as the four rank 
tensor
$$
R^{i}_{klm}(x)=\frac{\partial x^{i}}{\partial y^{s}}
\frac{\partial y^{r}}{\partial x^{k}}
\frac{\partial y^{q}}{\partial x^{l}} 
\frac{\partial y^{p}}{\partial x^{n}}
R^{i}_{klm}(y)
\eqno(160)
$$                                             
the curvature tensor for 
the coordinate system 
$(x)$ is also equal to 
zero
$$
R^{i}_{klm}(x)=0.  \eqno(161) 
$$                                                
Hence,  
for the coordinate system $(x)$ we get the following system of 
three equations
$$
\frac{\partial \Gamma_{m}}{\partial x^{l}}
-\frac{\partial \Gamma_{l}}{\partial x^{n}}
+\Gamma_{l}\Gamma_{m}-
\Gamma_{m}\Gamma_{l}=0
\eqno(162)
$$  
where $\Gamma_{m}(x)$  are matrices with components 
$$
\Gamma^{k}_{ij}=\frac{1}{2}g^{kl}(g_{il,j}+g_{jl,i}-g_{ij,l}). \eqno(163)
$$
We note that in our case the scalar curvature is equal to zero
$$
R=\sum^{3}_{i,k,l,m=1}g^{il}g^{km}R_{iklm}=0.  \eqno(164)
$$
 
Now the system (155) takes the form
$$
\frac{\partial^{2}y^{k}}{\partial x^{i}\partial x^{j}}=
\Gamma^{l}_{ij}(x)\frac{\partial y^{k}}{\partial x^{l}}.  \eqno(165)
$$ 

Let ${\bf r}=(y^{1},y^{2},y^{3}) ={\bf r}(x^{1},x^{2},x^{3})$ is the 
position vector and put $x^{1}=x, 
x^{2}=y, x^{3}=t$. Then as follows from (165) the position vector 
${\bf r}$ 
satisfies the following set of equations
$$
{\bf r}_{xx} = \Gamma^{1}_{11} {\bf r}_{x} + \Gamma^{2}_{11} {\bf r}_{y}+ 
\Gamma^{3}_{11} {\bf r}_{t}  \eqno(166a)
$$ 
$$
{\bf r}_{xy} = \Gamma^{1}_{12} {\bf r}_{x} + \Gamma^{2}_{12} {\bf r}_{y}+ 
\Gamma^{3}_{12} {\bf r}_{t}  \eqno(166b)
$$ 
$$
{\bf r}_{xt} = \Gamma^{1}_{13} {\bf r}_{x} + \Gamma^{2}_{13} {\bf r}_{y}+ 
\Gamma^{3}_{13} {\bf r}_{t}  \eqno(166c) $$
$$
{\bf r}_{yy} = \Gamma^{1}_{22} {\bf r}_{x} + \Gamma^{2}_{22} {\bf r}_{y}+ 
\Gamma^{3}_{22} {\bf r}_{t}  \eqno(166d) $$
$$
{\bf r}_{yt} = \Gamma^{1}_{23} {\bf r}_{x} + \Gamma^{2}_{23} {\bf r}_{y}+ 
\Gamma^{3}_{23} {\bf r}_{t}  \eqno(166e)
$$ 
$$
{\bf r}_{tt} = \Gamma^{1}_{33} {\bf r}_{x} + \Gamma^{2}_{33} {\bf r}_{y}+ 
\Gamma^{3}_{33} {\bf r}_{t}.  \eqno(166f)
$$
We can rewrite the equation (166) in the 
following form 
$$ 
Z_{x} = A_{1}Z, \quad Z_{y} = A_{2}Z, \quad Z_{t} = 
A_{3}Z  \eqno(167) 
$$
where
$$
Z=({\bf r}_{x}, {\bf r}_{y}, {\bf r}_{t})^{T}
\eqno(168)
$$ 
and
$$
A_{1} =
\left ( \begin{array}{ccc}
\Gamma^{1}_{11} & \Gamma^{2}_{11} & \Gamma^{3}_{11} \\
\Gamma^{1}_{12} & \Gamma^{2}_{12} & \Gamma^{3}_{12}\\
\Gamma^{1}_{13}           &\Gamma^{2}_{13}     & \Gamma^{3}_{13}
\end{array} \right)   \quad
A_{2} =
\left ( \begin{array}{ccc}
\Gamma^{1}_{12} & \Gamma^{2}_{12} & \Gamma^{3}_{12}\\
\Gamma^{1}_{22} & \Gamma^{2}_{22} & \Gamma^{3}_{22}\\
\Gamma^{1}_{23}         & \Gamma^{2}_{23}          & \Gamma^{3}_{23}
\end{array} \right)
$$
$$
A_{3} =
\left ( \begin{array}{ccc}
\Gamma^{1}_{13} & \Gamma^{2}_{13} & \Gamma^{3}_{13} \\
\Gamma^{1}_{23} & \Gamma^{2}_{23} & \Gamma^{3}_{23} \\
\Gamma^{1}_{33}          & \Gamma^{2}_{33}  & \Gamma^{3}_{33}
\end{array} \right). \eqno(169)
$$
The compatibility condition of these equations are given by
$$
\frac{\partial A_{i}}{\partial x^{j}}-\frac{\partial A_{j}}{\partial 
x^{i}}  + [A_{i},A_{j}] = 0.  \eqno(170)
$$

For our further work it is convenient use the triad of unit vectors. 
We introduce these vectors by the way
$$
{\bf e}_{1} = \frac{{\bf r}_{x}}{H_{1}}, \quad    {\bf e}_{2} = \frac{{\bf r}_{y}}{H_{2}} +c_{1}{\bf r}_{x}+c_{2}{\bf r}_{t},
\quad {\bf  e}_{3} = {\bf e}_{1} \wedge {\bf  e}_{2}.    \eqno(171)
$$
The explicit forms of  $c_{1}, c_{2}$  given in [2] and
$$
H_{1}=\mid{\bf r}_{x}\mid, \quad H_{2}=\mid{\bf r}_{y}\mid, \quad H_{3}=
\mid{\bf r}_{t}\mid.  \eqno(172)
$$
The equations (156) for ${\bf e}_{k}$ take the form 
$$
\left ( \begin{array}{c}
{\bf e}_{1} \\
{\bf e}_{2} \\
{\bf e}_{3}
\end{array} \right)_{x}= B_{1}
\left ( \begin{array}{ccc}
{\bf e}_{1} \\
{\bf e}_{2} \\
{\bf e}_{3}
\end{array} \right), \quad
\left ( \begin{array}{ccc}
{\bf e}_{1} \\
{\bf e}_{2} \\
{\bf e}_{3}
\end{array} \right)_{y}= B_{2}
\left ( \begin{array}{ccc}
{\bf e}_{1} \\
{\bf e}_{2} \\
{\bf e}_{3}
\end{array} \right)
$$
$$
\left ( \begin{array}{ccc}
{\bf e}_{1} \\
{\bf e}_{2} \\
{\bf e}_{3}
\end{array} \right)_{t}= B_{3}
\left ( \begin{array}{ccc}
{\bf e}_{1} \\
{\bf e}_{2} \\
{\bf e}_{3}
\end{array} \right)
\eqno(173)
$$
with
$$
B_{1} =
\left ( \begin{array}{ccc}
0             & k     &  -\sigma \\
-\beta k      & 0     & \tau  \\
\beta\sigma        & -\tau & 0
\end{array} \right) ,
B_{2}=
\left ( \begin{array}{ccc}
0            & m_{3}  & -m_{2} \\
-\beta m_{3} & 0      & m_{1} \\
\beta m_{2}  & -m_{1} & 0
\end{array} \right)
$$
$$
B_{3}=
\left ( \begin{array}{ccc}
0       & \omega_{3}  & -\omega_{2} \\
-\beta\omega_{3} & 0      & \omega_{1} \\
\beta\omega_{2}  & -\omega_{1} & 0
\end{array} \right) \eqno(174)
$$
where $k, \tau, \sigma, m_{i}, \omega_{i}$ are some real functions the 
explicit forms of which given in [?], $\beta={\bf e}_{1}^{2}=\pm 1, 
{\bf e}^{2}_{2}={\bf e}^{2}_{3}=1$. 
Again from the integrability condition of these equations we obtain the following set of equations
$$
\frac{\partial B_{i}}{\partial x^{j}}-\frac{\partial B_{j}}{\partial
x^{i}}  + [B_{i},B_{j}] = 0.  \eqno(175)
$$     

Many integrable systems in 2+1 dimensions are exact reductions of the 
equation (175) (see, e.g., [53]).

\section{Integrable reductions}

Now to find out particular integrable 
reductions of three-dimensional Riemann space, as in [49-51] ,
we will use multidimensional integrable spin systems (MISSs). To 
this end, we assume that 
$$
{\bf e}_{1} \equiv {\bf S} \eqno(176)
$$
where ${\bf S}=(S_{1}, S_{2}, S_{3})$ is the spin vector, ${\bf 
S}^{2}=\beta=\pm 1$. So, the vector ${\bf e}_{1}$ satisfies the some 
given MISS.  Some comments on MISSs 
are in order. At present there exist several MISSs (see, e.g., Appendix). 
They play important 
role both in mathematics and physics.  In this note, to find out a 
integrable 
case of three-dimensional Riemann space we will use the 
MISS - the Ishimori equation (IE).

\subsection{The Ishimori equation}
The IE has the form 
$$
{\bf S}_t = {\bf S} \wedge ({\bf S}_{xx}+\alpha^{2} S_{yy})
+u_{y}{\bf S}_x+u_{x}{\bf S}_y  \eqno(177a)
$$
$$
u_{xx}-\alpha^{2}u_{yy}=-2\alpha^{2}{\bf S}({\bf S}_x \wedge {\bf S}_y).
\eqno(177b)
$$
The IE is integrable by Inverse Scattering Transform (IST) [42]. 
In this case,  we get 
$$
m_{1}=\partial_{x}^{-1}[\tau_{y}-\frac{1}{2\alpha^2}M^{\prime}_2 u],\quad
m_{2}=-\frac{1}{2\alpha^2 k}M^{\prime}_2 u
$$
$$
m_{3}=\partial_{x}^{-1}[k_y +\frac{\tau}{2\alpha^2 k}M_2 u],
\quad M^{\prime}_{2}u=\alpha^{2}u_{yy}-u_{xx}  \eqno(178)
$$
and 
$$
\omega_{2}= -(k_{x}+\sigma \tau)-
\alpha^{2}(m_{3y}+m_{2}m_{1})
+m_{2}u_{x}+\sigma u_{y}
$$
$$
\omega_{3}= (\sigma_{x}-k \tau)+
\alpha^{2}(m_{2y}-m_{3}m_{1})
+k u_{y}+m_{3}u_{x}
$$
$$
\omega_{1} = \frac{1}{k}[\sigma_{t}-\omega_{2x}+\tau\omega_{3}]. \eqno(179)
$$
                                                                    
Thus we expressed the functions $m_{k}$ and $\omega_{k}$ by the three 
functions $k,\tau, \sigma$ and their derivatives. This means that we 
identified the equations (161) and (184) which define the geometry of 
Riemann space with the given MISS - the IE (177). In 
turn it means 
that we given off the integrable case of the three-dimensional Riemann 
space. Hence arises 
the natural question: how construct the integrable three-dimensional 
Riemann space using the MISS or that the same thing how find $g_{ij}$?.   
The answer as follows. Let 
$$
{\bf r}_{x}^{2}=H^{2}_{1}=\beta =\pm 1.  \eqno(180)
$$
Then we have
that 
$$
{\bf r}=\partial^{-1}_{x}{\bf S}+{\bf r}_{0}(y,t)   \eqno(181)
$$
where $\partial^{-1}_{x}=\int^{x}_{-\infty}dx$. For simplicity, 
we put ${\bf r}_{0}=0$. Now we 
can express the coefficients of the metric (143) by ${\bf S}$. As shown in 
[50], for the (2+1)-dimensional case, we have $$
g_{11} = {\bf S}^{2}=\pm 1, \quad g_{12}={\bf S}\cdot \partial^{-1}_{x}{\bf 
S}_{y}, \quad  g_{13}={\bf S}\cdot\partial^{-1}_{x}{\bf S}_{t}
$$
$$
g_{22}=(\partial^{-1}_{x}{\bf S}_{y})^{2},
\quad g_{23}=(\partial^{-1}_{x}{\bf S}_{y})\cdot (\partial^{-1}_{x}{\bf 
S}_{t}), \quad g_{33}=(\partial^{-1}_{x}{\bf S}_{t})^{2}.
\eqno(182)
$$

\subsection{The Davey-Stewartson  
equation}
It is well known from the soliton theory that between integrable spin 
systems and NLS-type equations take place the so-called gauge and/or 
L-equivalences [40, 42]. From this fact and from the identification the 
three-dimensional Riemann space and the MISS (177) in the previous 
subsection follows that there exist some connections with NLS-type 
equations. We show it in this subsection. Let us  we introduce two complex 
functions 
$q, p$ as $$ q = a_{1}e^{ib_{1}}, \quad p=a_{2}e^{ib_{2}}  \eqno(183)
$$
where $a_{j}, b_{j}$ are real functions. Let $a_{k}, b_{k}$ have the form
$$
a_{1}^2 =\frac{|a|^2}{|b|^2}\{\frac{k^2}{4}
+\frac{|\alpha|^2}{4}(m_3^2 +m_2^2)-\frac{1}{2}\alpha_{R}km_3-
\frac{1}{2}\alpha_{I}km_2\}
  \eqno(184a)
$$
$$
b_{1} =\partial_{x}^{-1}\{-\frac{\gamma_1}{2ia_1^{\prime^{2}}}-(\bar 
A-A+D-\bar D)\}  \eqno(184b)
$$ 
$$
a_{2}^2 =\frac{|b|^2}{|a|^2}\{\frac{k^2}{4}
+\frac{|\alpha|^2}{4}(m_3^2 
+m_2^2)+\frac{1}{2}\alpha_{R}km_3-\frac{1}{2}\alpha_{I}km_2\}
  \eqno(184c)
$$
$$
b_{2} =\partial_{x}^{-1}\{-\frac{\gamma_2}{2ia_2^{\prime^{2}}}-(A-\bar 
A+\bar D-D)  \eqno(184d) 
$$
where
$$
\gamma_1=i\{\frac{1}{2}k^{2}\tau+\frac{|\alpha|^2}{2}(m_3km_1+m_2k_y)-
$$
$$
\frac{1}{2}\alpha_{R}[k^{2}m_1+m_3k\tau+
m_2k_x]+\frac{1}{2}\alpha_{I}[k(2k_y-m_{3x})-
k_x m_3]\} \eqno(185a)
$$
$$
\gamma_2=-i\{\frac{1}{2}k^{2}\tau+
\frac{|\alpha|^2}{2}(m_3km_1+m_2k_y)+
$$
$$
\frac{1}{2}\alpha_{R}(k^{2}m_1+m_3k\tau+
m_2k_x)+\frac{1}{2}\alpha_{I}[k(2k_y-m_{3x})-
k_x m_3]\}. \eqno(185b)
$$
Here $\alpha=\alpha_{R}+i\alpha_{I}$. In this case, $q,p$
satisfy the DS equation [?] 
$$
iq_t+q_{xx}+\alpha^{2}q_{yy}+vq=0 \eqno(186a)
$$
$$
ip_t-p_{xx}-\alpha^{2}p_{yy}-vp=0 \eqno(186b)
$$
$$
\alpha^{2}v_{yy}--v_{xx}=-2[\alpha^{2}(pq)_{yy}+(pq)_{xx}]. \eqno(186c)
$$

\section{Diagonal metrics}

Now we consider the case when the metric has the diagonal form, i.e.
$$
ds^{2}=\epsilon_{1}H_{1}^{2}dx^{2} 
+\epsilon_{2}H^{2}_{2}dy^{2}+\epsilon_{3}H_{3}^{2}dt^{2}  \eqno(187) 
$$
where $\epsilon_{i}=\pm 1.$ In this note we consider the case when 
$\epsilon_{i}=+1$.  In this case, the Christoffel symbols take the form
$$
\Gamma^{1}_{11}=\frac{H_{1x}}{H_{1}}=\beta_{11}, \quad   
\Gamma^{2}_{11}=-\frac{H_{1}H_{1y}}{H^{2}_{2}}=
-\frac{H_{1}}{H_{2}}\beta_{21}, \quad   
\Gamma^{3}_{11}=-\frac{H_{1}H_{1t}}{H_{3}^{2}}=-\frac{H_{1}}{H_{3}}\beta_{31} 
$$
$$
\Gamma^{1}_{12}=\frac{H_{1y}}{H_{1}}=\frac{H_{2}}{H_{1}}\beta_{21}, \quad
\Gamma^{2}_{12}=\frac{H_{2x}}{H_{2}}=\frac{H_{1}}{H_{2}}\beta_{12}, \quad 
\Gamma^{3}_{12}=0 
$$
$$
\Gamma^{1}_{13}=\frac{H_{1t}}{H_{1}}=\frac{H_{3}}{H_{1}}\beta_{31}, \quad
\Gamma^{2}_{13}=0, \quad  
\Gamma^{3}_{13}=\frac{H_{3x}}{H_{3}}=\frac{H_{1}}{H_{3}}\beta_{13} 
$$ 
$$
\Gamma^{1}_{22}=-\frac{H_{2}H_{2x}}{H_{1}^{2}}=
-\frac{H_{2}}{H_{1}}\beta_{12},\quad  
\Gamma^{2}_{22}=\frac{H_{2y}}{H_{2}}=\beta_{22}, \quad 
\Gamma^{3}_{22}=-\frac{H_{2}H_{2t}}{H^{2}_{3}}=
-\frac{H_{2}}{H_{3}}\beta_{32}  
$$
$$
\Gamma^{1}_{23}=0, \quad
\Gamma^{2}_{23}=\frac{H_{2t}}{H_{2}}=\frac{H_{3}}{H_{2}}\beta_{32},
\quad  \Gamma^{3}_{23}= 
\frac{H_{3y}}{H_{3}}=\frac{H_{2}}{H_{3}}\beta_{23} 
$$
$$
\Gamma^{1}_{33}=-\frac{H_{3}H_{3x}}{H_{1}^{2}}=-\frac{H_{3}}{H_{1}}\beta_{13}, \quad 
\Gamma^{2}_{33}=-\frac{H_{3}H_{3y}}{H_{2}^{2}}=-\frac{H_{3}}{H_{2}}\beta_{23}, \quad  
\Gamma^{3}_{33}=\frac{H_{3t}}{H_{3}}=\beta_{33} 
$$
or
$$
\Gamma^{k}_{ij}=0 
\quad i\neq 
j\neq k 
\eqno(188a)
$$
$$
\Gamma^{i}_{il}=\frac{H_{i,l}}{H_{i}}=\frac{H_{l}}{H_{i}}\beta_{li}  
\eqno(188b)
$$
$$
\Gamma^{i}_{ll}=-\frac{H_{l}H_{li}}{H_{i}^{2}}=-\frac{H_{l}}{H_{i}}\beta_{il}, 
\quad i\neq l. \eqno(188c) 
$$
Here
$$
\beta_{ik}=\frac{H_{k,i}}{H_{i}} \eqno(189)
$$
are the so-called rotation coefficients. 
      In this case we get 
$$
\tau=m_{2}=\omega_{3}=0  
\eqno(190) 
$$ 
and the matrices $B_{i}$ take the 
form 
$$
B_{1} =
\left ( \begin{array}{ccc}
0             & -\beta_{21}     &  -\beta_{31} \\
\beta_{21}       & 0     &   0 \\
\beta_{31}        &    0 & 0
\end{array} \right),\quad
B_{2}=
\left ( \begin{array}{ccc}
0            & \beta_{12}  & 0 \\
-\beta_{12} & 0      & -\beta_{32}\\
0  & \beta_{32} & 0
\end{array} \right)
$$
$$
B_{3} =
\left ( \begin{array}{ccc}
0       & 0  & \beta_{13} \\
0 & 0      & \beta_{23} \\
-\beta_{13}  & -\beta_{23} & 0
\end{array} \right).\eqno(191)
$$ 
So we have
$$
\beta_{23x}=\beta_{13}\beta_{21}, \quad \beta_{32x}=\beta_{12}\beta_{31}
   \eqno(192a)
$$     
$$
\beta_{13y}=\beta_{12}\beta_{23}, \quad \beta_{31y}=\beta_{32}\beta_{21}
   \eqno(192b)
$$
$$
\beta_{12t}=\beta_{13}\beta_{32}, \quad \beta_{21t}=\beta_{23}\beta_{31}
   \eqno(192c)
$$
$$
\beta_{12x}+\beta_{21y}+\beta_{31}\beta_{32}=0
\eqno(192d)
$$
$$
\beta_{13x}+\beta_{31t}+\beta_{21}\beta_{23}=0
   \eqno(192e)
$$
$$
\beta_{23y}+\beta_{32t}+\beta_{12}\beta_{13}=0
   \eqno(192f)
$$
or
$$
\frac{\partial \beta_{ij}}{\partial x^{k}}=\beta_{ik}\beta_{kj}, \quad 
i\neq j \neq k  \eqno(192g) $$
$$
\frac{\partial \beta_{ij}}{\partial 
x^{i}}+\frac{\partial \beta_{ji}}{\partial x^{j}}
+\sum^{3}_{m\neq i,j} \beta_{mi}\beta_{mj}=0, \quad i\neq j  \eqno(192h)
$$                                      
This nonlinear system is the famous Lame equation and well-known in the 
theory of 3-orthogonal 
coordinates. The problem of description of curvilinear orthogonal 
coordinate systems in a a (pseudo-) Euclidean space is a classical 
problem of differential geometry. It was studied in detail and mainly 
solved in the beginning of the 20th century. Locally, such coordinate 
systems are determinated by $\frac{n(n-1)}{2}$ arbitrary functions of two 
variables.  This problem in 
some sense is equivalent to the problem of description of diagonal flat 
metrics, that is, flat metrics $g_{ij}(x)=f_{i}(x)\delta_{ij}$. 
We mention that the Lame equation describing curvilinear orthogonal 
coordinate systems can be  integrated by 
the IST [33] (see also an algebraic-geometric approach in [34]). 
Now we would 
like consider some particular cases (see, also, e.g., [26]).

\subsection{$H_{1}=H_{2}=H,\quad H_{3}=1$}
Let $H=e^{\psi}$. We get the following set of equations
$$
\psi_{xx}+\psi_{yy}+\psi^{2}_{t}e^{2\psi}=0 \eqno(193a)
$$
$$
\psi_{tx}=\psi_{ty}=0, \quad \psi_{tt}+\psi^{2}_{t}=0. \eqno(193b)
$$
From (159) we obtain
$\psi_{t}=\frac{1}{C-t}$. So Eq. (193) reduced to the equation
$$
\psi_{xx}+\psi_{yy}+\frac{1}{C-t}e^{2\psi}=0. \eqno(194)
$$ 
\subsection{$H_{1}=\cos \theta, 
\quad H_{2}=\sin \theta, \quad H_{3}=1$}
In this case the corresponding equation has the form
$$
(\theta_{t}\cos \theta)_{x}=-\theta_{x}\theta_{t}\sin \theta, \quad 
(\theta_{t}\sin \theta)_{y}=\theta_{y}\theta_{t}\cos \theta  \eqno(195a)
$$
$$
\theta_{xt}=\theta_{yt}=0, \quad 
\theta_{xx}-\theta_{yy}-\theta_{t}^{2}\sin\theta\cos\theta=0 \eqno(195b)
$$
$$
(\theta_{t}\sin \theta)_{t}=(\theta_{t}\cos \theta)_{t}=0. \eqno(195c)
$$
 
\subsection{$H_{1}=\cos 
\theta, \quad H_{2}=\sin \theta,\quad H_{3}=\theta_{t}$}
In this case the corresponding equation looks like
$$
(\frac{\theta_{ty}}{\sin \theta})_{x}=-\frac{\theta_{xt}\theta_{y}}{\cos 
\theta}  \eqno(196a)
$$
$$
(\frac{\theta_{xt}}{\cos \theta})_{y}=\frac{\theta_{x}\theta_{ty}}{\sin 
\theta} \eqno(196b)
$$
$$
\theta_{xx}-\theta_{yy}-\sin \theta \cos \theta=0 \eqno(196c)
$$
$$
(\frac{\theta_{xt}}{\cos \theta})_{y}-(\sin 
\theta)_{t}-\frac{\theta_{y}\theta_{ty}}{\sin \theta}=0 \eqno(196d)
$$
$$
(\frac{\theta_{ty}}{\sin \theta})_{y}+(\cos 
\theta)_{t}+\frac{\theta_{x}\theta_{xt}}{\cos \theta}=0. \eqno(196e)
$$

\subsection{$H_{1}=e^{\psi}, \quad H_{2}=e^{\psi},\quad 
H_{3}=\psi_{t}$}
In this case we have
$$
(\psi_{ty}e^{-\psi})_{x}=\psi_{tx}\psi_{y}e^{-\psi}, \quad 
(\psi_{tx}e^{-\psi})_{y}=\psi_{ty}\psi_{x}e^{-\psi} \eqno(197a)
$$
$$
\psi_{xx}+\psi_{yy}+e^{2\psi}=0  \eqno(197b)
$$
$$
(\psi_{tx}e^{-\psi})_{x}+\psi_{t}e^{\psi}+\psi_{y}\psi_{ty}e^{-\psi}=0,
\quad
(\psi_{ty}e^{-\psi})_{y}+\psi_{t}e^{\psi}+\psi_{x}\psi_{tx}e^{-\psi}=0.
\eqno(197c)
$$ 
\subsection{$H_{1}=H_{2}=H_{3}= H^{2}$}
In this choose we get ($H=e^{\psi}$)
$$
\psi_{xy}=\psi_{x}\psi_{y}, \quad 
\psi_{xt}=\psi_{x}\psi_{t}, \quad
\psi_{yt}=\psi_{t}\psi_{y}  \eqno(198a)
$$
$$
\psi_{xx}+\psi_{yy}+4\psi^{2}_{t}=0, \quad
\psi_{xx}+\psi_{tt}+4\psi^{2}_{y}=0, 
\quad \psi_{tt}+\psi_{yy}+4\psi^{2}_{x}=0. \eqno(198b)
$$
                 
\section{Connections with the other equations}
It is remarkable that the equation (175) [=(162)=(170)=(213)] is related 
with the some well-known equations. In this section we 
present some of these connections.
\subsection{Equation 
(175) and the Bogomolny equation} Consider the Bogomolny 
equation (BE) [31] $$
\Phi_{t}+[\Phi,B_{3}]+ B_{1y} - B_{2x} + [B_{1},B_{2}] = 0 \eqno(199a)
$$
$$
\Phi_{y}+[\Phi,B_{2}]+B_{3x}-B_{1t}+[B_{3},B_{1}]=0   \eqno(199b)
$$
$$
\Phi_{x}+[\Phi,B_{1}]+B_{2t}-B_{3y}+[B_{2},B_{3}]=0. \eqno(199c)
$$
This equation is integrable and play important role in the field theories 
in particular in the theory of monopols. 
The set of equations (175) is the particular case of the BE. In fact, 
as $\Phi = 0$ from (199) we obtain the system (175). 

\subsection{Equation (175) and the Self-Dual 
Yang-Mills  equation} 

Equation (175) is exact reduction of the SO(3)-Self-Dual 
Yang-Mills equation (SDYME)
$$
F_{\alpha\beta}=0, \quad F_{\bar\alpha\bar\beta}=0,
\quad
F_{\alpha\bar\alpha}+F_{\beta\bar\beta}=0  \eqno(200)
$$
Here
$$
F_{\mu\nu}=\frac{\partial A_{\nu}}{\partial x_{\mu}}-
\frac{\partial A_{\mu}}{\partial x_{\nu}}+[A_{\mu},A_{\nu}] \eqno(201)
$$
and
$$
\frac{\partial}{\partial x_{\alpha}}=\frac{\partial}{\partial z}-
i\frac{\partial}{ \partial t}, \quad
\frac{\partial}{\partial x_{\bar\alpha}}=\frac{\partial}{\partial z}+
i\frac{\partial}{ \partial t}, \quad
\frac{\partial}{\partial x_{\beta}}=\frac{\partial}{\partial x}-
i\frac{\partial}{ \partial y}
$$
$$
\frac{\partial}{\partial x_{\bar\beta}}=\frac{\partial}{\partial x}+
i\frac{\partial}{ \partial y}. \eqno(202)
$$
In fact, if in the SDYME (200) we take
$$
A_{\alpha}= -iB_{3}, \quad A_{\bar\alpha}= iB_{3}, \quad
A_{\beta}= B_{1}-iB_{2}, \quad A_{\beta}= B_{1}+iB_{2} \eqno(203)
$$
and if $B_{k}$ are independent of $z$, then the SDYME (200)
reduces to  the  equation (175).
As known that the LR of the SDYME has the form [41, 43]
$$
(\partial_{\alpha}+\lambda\partial_{\bar\beta})\Psi=(A_{\alpha}+\lambda 
A_{\bar \beta})\Psi, 
\quad
(\partial_{\beta}-\lambda\partial_{\bar\alpha})\Psi=(A_{\beta}-
\lambda A_{\bar\alpha})\Psi \eqno(204)
$$
where $\lambda$ is the spectral parameter  satisfing the following
set of the equations
$$
\lambda_{\beta}=\lambda\lambda_{\bar\alpha}, \quad
\lambda_{\alpha}=-\lambda\lambda_{\bar\beta}. \eqno(205)
$$
Apropos, the simplest solution of this set has may be the following
form 
$$
\lambda 
=\frac{a_{1}x_{\bar\alpha}+a_{2}x_{\bar\beta}+a_{3}}{a_{2}x_{\alpha}-
a_{1}x_{\beta}+a_{4}}, \quad a_{j}=consts. \eqno(206)
$$
From (204) we obtain the LR of the equation (175)
$$
(-i\partial_{t}+\lambda\partial_{\bar\beta})\Psi=[-iB_{3}+\lambda 
(B_{1}+iB_{2})]\Psi \eqno(207a)
$$
$$
(\partial_{\beta}-i\lambda\partial_{t})\Psi=[(B_{1}-iB_{2})-i\lambda 
B_{3}]\Psi.
 \eqno(207b)
$$     

\subsection{Equation (175) and the Chern-Simons equation}

Consider the action of the Chern-Simons (CS) theory [44]
$$
S[J]=\frac{k}{4\pi}\int_{M}tr(J\wedge dJ+\frac{2}{3}J\wedge J\wedge J)
\eqno(208)
$$
where $J$ is a 1-form gauge connection with values in the Lie algebra 
$\hat g$ of a (compact or noncompact) non-Abelian simple Lie group $\hat 
G$ on an oriented closed 3-dimensional manifold $M$, $k$ is the coupling 
constant. The classical equation of motion is the zero-curvature condition
$$ 
dJ+J\wedge J=0.  \eqno(209) $$ 
Let the 1-form $J$ has the form
$$
J=B_{1}dx+B_{2}dy+B_{3}dt.  \eqno(210)
$$
As shown in [44], subtituting the (210) into (209) we obtain the equation 
(175).  Note that from this fact and from the results of 
the subsection 5.2 follows that the CS - equation of motion (209) is exact 
reduction of the SDYM equation (200).

\subsection{Equation (175) as some generalization of the Lame equation}
Let us the matrices $B_{i}$ (174) we rewrite in the form
$$
B_{1} =
\left ( \begin{array}{ccc}
0             & -\beta_{21}     &  -\beta_{31} \\
\beta_{21}       & 0     &   \tau \\
\beta_{31}        & -\tau & 0
\end{array} \right),\quad
B_{2}=
\left ( \begin{array}{ccc}
0            & \beta_{12}  & -m_{2} \\
-\beta_{12} & 0      & -\beta_{32}\\
m_{2}  & \beta_{32} & 0
\end{array} \right)
$$
$$
B_{3} =
\left ( \begin{array}{ccc}
0       & \omega_{3}  & \beta_{13} \\
-\omega_{3} & 0      & \beta_{23} \\
-\beta_{13}  & -\beta_{23} & 0
\end{array} \right).\eqno(211)
$$

Then the equation (175) in elements takes the form
$$
\beta_{23x}-\tau_{t}=\beta_{13}\beta_{21}-\omega_{3}\beta_{31} \eqno(212a)
$$
$$
 \beta_{32x}+\tau_{y}=\beta_{12}\beta_{31}+m_{2}\beta_{21}
   \eqno(212b)
$$
$$
\beta_{13y}+m_{2t}=\beta_{12}\beta_{23}+\omega_{3}\beta_{32}
\eqno(212c)
$$
$$
 \beta_{31y}-m_{2x}=\beta_{32}\beta_{21}-\tau\beta_{12}
   \eqno(212d)
$$
$$
\beta_{12t}-\omega_{3y}=\beta_{13}\beta_{32}-m_{3}\beta_{23}
\eqno(212e)
$$
$$
 \beta_{21t}+\omega_{3x}=\beta_{23}\beta_{31}+\tau\beta_{13}   \eqno(212f)
$$
$$
\beta_{12x}+\beta_{21y}+\beta_{31}\beta_{32}+\tau m_{2}=0
\eqno(212g)
$$
$$
\beta_{13x}+\beta_{31t}+\beta_{21}\beta_{23}+\tau\omega_{3}=0
   \eqno(212h)
$$
$$
\beta_{23y}+\beta_{32t}+\beta_{12}\beta_{13}+m_{2}\omega_{3}=0
   \eqno(212i)
$$
Hence as 
$\tau=m_{2}=\omega_{3}=0 $
 we obtain the Lame equation (192). So, the equation (175) is one of the 
generalizations of the Lame equation.

\section{On Lax representation of the equation (175)}

As follows from the results of  the  section 3, Equation (175) can  
admits several integrable reductions. At the same time, the results of 
the subsections 5.1-5.2 show that the equation (175) is integrable 
may be and in general case.  At least, it admits the LR of the form (207) 
and/or of the following form (see, e.g., [38]) 
$$
\Phi_{x}=U_{1}\Phi, \quad  \Phi_{y}=U_{2}\Phi, \quad \Phi_{t}=U_{3}\Phi 
\eqno(213) 
$$
with
$$
U_{1}=\frac{1}{2} 
\left ( \begin{array}{cc}
i\tau          & -(k+i\sigma) \\
k-i\sigma       & -i\tau
\end{array} \right),\quad
U_{2}=
\frac{1}{2}\left ( \begin{array}{cc}
im_{1}       & -(m_{3}+im_{2}) \\
m_{3}-im_{2} &  -im_{1}
\end{array} \right)
$$
$$
U_{3}=
\frac{1}{2}\left ( \begin{array}{cc}
i\omega_{1}       & -(\omega_{3}+i\omega_{2}) \\
\omega_{3}-i\omega_{2} &  -i\omega_{1}
\end{array} \right).  \eqno(214)
$$
Systems of this type were first studied by Zakharov and Shabat [35]. The 
integrability conditions on this system of overdetermined equations 
(211), require that 
$$
U_{i,j}-U_{j,i}+[U_{i}, U_{j}]=0. \eqno(215)
$$
Many (and perhaps all) integrable systems in 2+1 dimensions have the LR of 
the form (211). 
In our case, the IE (177) and the DS equation (186) have also the LR of 
the form (211) with 
the functions $m_{i}, \omega_{i}$ given by (123) and (124). On the other 
hand, it is well-known that  for example  the DS equation (186) has the 
following LR of the standard form
$$
\alpha\Psi_{y}=\sigma_{3}\Psi_{x}+Q\Psi,  \quad 
Q=\left ( \begin{array}{cc}
0   & q\\
p   &  0
\end{array} \right)  
 \eqno(216a)
$$
$$
\Psi_{t}=2i\sigma_{3}\Psi_{xx}+2iQ\Psi_{x}+
\left ( \begin{array}{cc}
c_{11}       & iq_{x}+i\alpha q_{y} \\
ip_{x}-i\alpha p_{y} &  c_{22}
\end{array} \right)\Psi  \eqno(216b)
$$
with
$$
c_{11x}-\alpha c_{11y}=i[(pq)_{x}+\alpha(pq)_{y}], \quad 
c_{22x}+\alpha c_{22y}=-i[(pq)_{x}-\alpha(pq)_{y}]. \eqno(217)
$$

Hence arises the natural question: how connected the 
both LR for one and the same integrable systems (in our case 
for the DS equation)?.                    
In fact, these two LR are related by the gauge transformation [50-51]
$$
\Phi=g \Psi  \eqno(218)
$$
where $\Phi$ and $\Psi$ are some solutions of the equations (222) and (111), 
respectively, while $g$ is the some matrix. 

\section{Conclusion}
In this note we have studied the relation between integrable systems in 
2+1 dimensions and 3-dimensional Riemann spaces. 
We have shown that in this 
geometrical setting certain typical structures of the completely 
integrable (2+1)-dimensional systems arise. 
To find out the integrable cases of the 3-dimensional Riemann space as an 
examples we used the Ishimori and Davey-Stewartson equations. The 
connections of the equations characterizing of the 3-dimensional Riemann 
space with the other known equations such as the Bogomolny and Self-Dual 
Yang-Mills equations are considered. Such connection with the 
Chern-Simons equation of motion was established in [?].   Of course our 
approach needs further developments.  Finally we note that the details 
of some calculations were given in [50-51].

\section{Acknowledgments}

This work was
partially supported by INTAS (grant 99-1782). RM would like to thanks to  V.S.Dryuma, M.Gurses, B.G. Konopelchenko, 
D.Levi, L.Martina and G.Soliani for very helpful discussions and 
especially 
D.Levi for the financial support and kind hospitality.
 He is grateful to the EINSTEIN Consortium of Lecce University and the Department of 
Mathematics 
of Bilkent University for their financial supports and warm 
hospitality.  

\section{Appendix: MISSs}

At present there exist several MISSs. Here we will give some of them.

i) {\it The Myrzakulov I (M-I) equation} 
The simplest example of MISS is the  M-I (Remark: about our notations please 
see e.g., ref. [38]) equation looks like   
$$
{\bf S}_{t}=({\bf S}\wedge {\bf S}_{y} +u{\bf S})_{x}  \eqno(219a)
$$
$$
u_{x}=-{\bf S}\cdot ({\bf S}_{x}\wedge {\bf S}_{y}) \eqno(219b)
$$

ii) {\it The Myrzakulov VIII (M-VIII) equation}. 
The M-VIII equation is one of simplest MISSs in 2+1 dimensions and reads 
as [52] 
$$
{\bf S}_{t}={\bf S}\wedge {\bf S}_{xx} +u{\bf S}_{x}  \eqno(220a)
$$
$$
u_{x}+u_{y}+{\bf S}\cdot ({\bf S}_{x}\wedge {\bf S}_{y})=0 \eqno(220b)
$$

iii) {\it The Ishimori equation}. The famous Ishimori equation has the 
form 
$$
{\bf S}_{t}={\bf S}\wedge ({\bf S}_{xx} + \alpha^{2}{\bf S}_{yy}) 
+u_{x}{\bf S}_{y}+u_{y}{\bf S}_{x}  \eqno(221a) $$
$$
u_{xx}-\alpha^{2}u_{yy}=
-2\alpha^{2}{\bf S}\cdot ({\bf S}_{x}\wedge {\bf S}_{y}) \eqno(221b) 
$$

iv) {\it The Myrzakulov IX (M-IX) equation}. This equation reads as [52]
$$
{\bf S}_{t}={\bf S}\wedge M_{1}{\bf S}-iA_{1}{\bf S}_{y}-iA_{2}{\bf 
S}_{x}  \eqno(222a)
$$
$$
M_{2}u =2\alpha^{2}{\bf S}\cdot ({\bf S}_{x}\wedge {\bf S}_{y}) \eqno(222b)
$$ 
Here  $M_{i}, A_{i}$ have the forms
$$
M_{1}=\alpha^{2}\frac{\partial^{2}}{\partial y^{2}} + 
4\alpha(b-a)\frac{\partial^{2}}{\partial x\partial 
y}+4(a^{2}-2ab-b)\frac{\partial^{2}}{\partial x^{2}}
$$
$$
M_{2}=\alpha^{2}\frac{\partial^{2}}{\partial y^{2}} -
2\alpha(2a+1)\frac{\partial^{2}}{\partial x\partial
y}+4a(a+1)\frac{\partial^{2}}{\partial x^{2}}
$$          
$$
A_{1}=i\{\alpha(2b+1)u_{y}-2(2ab+a+b)u_{x}\}
$$
$$
A_{2}=i\{4\alpha^{-1}(2a^{2}b+a^{2}+2ab+b)u_{x}-2(2ab+a+b)u_{y}\}
$$
The M-IX equation contains several particular integrable cases: a) the 
M-VIII equation (218) as $a=b=-1$; b) the Ishimori equation (177) 
as $a=b=-\frac{1}{2}$; c) the M-XXXIV equation 
as $a=b=-1, y=t$ [52]
$$
{\bf S}_{t}={\bf S}\wedge {\bf S}_{xx}+u{\bf S}_{x}  \eqno(223a) $$
$$
u_{t}+u_{x}+\frac{1}{2} ({\bf S}^{2}_{x})^{2}=0 
\eqno(223b) 
$$
and so on.  The M-XXXIV equation (221) describe the 
nonlinear dynamics of compressible magnets [45]. It is the first (and, to 
the best of our knowledge, at present the unique) example 
of integrable spin system governing the nonlinear interactions of spin 
(${\bf S})$ and lattice $(u)$ subsystems in 1+1 dimensions.

v) {\it The Myrzakulov XX (M-XX) equation}. This equation reads as [52]
$$
{\bf S}_{t}={\bf S}\wedge \{(b+1){\bf S}_{xx}-b{\bf S}_{yy}\}
+bu_{y}{\bf S}_{y}+(b+1)u_{x}{\bf S}_{x}  \eqno(224a)
$$
$$
u_{xy} ={\bf S}\cdot ({\bf S}_{x}\wedge {\bf S}_{y}) \eqno(224b)
$$   

vi) {\it The (2+1)-dimensional Myrzakulov 0 (M-0) equation}. The 
(2+1)-dimensional M-0  equation [52]
$$ {\bf S}_{t}=c_{1}{\bf
S}_{x}+c_{2}{\bf S}_{y}  \eqno(223) 
$$
is in 
general not integrable but admits integrable reductions. For example, the 
following case of the (2+1)-dimensional M-0 equation is integrable 
$$ {\bf S}_{t}=
-\frac{\beta_{13}}{\beta_{31}}
{\bf S}_{x}
-\frac{\beta_{13}}{\beta_{12}\beta_{31}}
{\bf S}_{y}  \eqno(226a) 
$$
$$
\frac{\partial \beta_{ij}}{\partial x^{k}}=\beta_{ik}\beta_{kj}  
\eqno(226b)
$$
$$
\frac{\partial \beta_{ij}}{\partial
x^{i}}+\frac{\partial \beta_{ji}}{\partial x^{j}}
+\sum^{3}_{m\neq i,j} \beta_{mi}\beta_{mj}=0  \eqno(226c)
$$
and so on.   All of these MISSs  are some 
(2+1)-dimensional integrable extensions of the isotropic Landau-Lifshitz 
(LL) equation
$$
{\bf S}_{t}={\bf S}\wedge {\bf S}_{xx}   \eqno(227)
$$
and in 1+1 dimensions reduced to it.
Here we would 
like mention that there exist the other classes MISSs 
which are not multidimensional generalizations of the 
LL equation (227), for example , the M-II, M-III and  M-XXII equations and 
so on. 
Finally we note that all MISSs in 2+1 dimensions are the integrable 
particular cases of the M-0 equation (225).

\end{document}